\documentclass[12pt]{article}

\usepackage{amsmath,amssymb,amsthm,epsf,graphicx}

\newcommand{\n}{\noindent}

\newcommand{\new}{\newpage}
\newcommand{\ka}{\kappa}
\newcommand{\ed}{\end{document}}
\newcommand{\be}{\begin{equation}}
\newcommand{\ee}{\end{equation}}

\begin{document}
\begin{center}
\large{\textbf{\textbf{THERMODYNAMICS OF PHOTON GAS WITH AN INVARIANT ENERGY SCALE}}}\\
\end{center}
\begin{center}
Sudipta Das$^{a,}$\footnote{E-mail: sudipta.das\_r@isical.ac.in},
Dibakar Roychowdhury$^{b,}$\footnote{E-mail: dibakar\_nbu@yahoo.co.in} \\
$^a$Physics and Applied Mathematics Unit, Indian Statistical
Institute\\ 203 B. T. Road, Kolkata 700108, India \\
$^b$Department of Physics, University of North Bengal \\ Siliguri
734013, West Bengal, India
\end{center}\vspace{1cm}

\begin{center}
{\textbf{Abstract}}
\end{center}

\n Quantum Gravity framework motivates us to find new theories in
which an observer independent finite energy upper bound
(preferably Planck Energy) exists. We have studied the
modifications in the thermodynamical properties of a photon gas in
such a scenario where we have an invariant energy scale. We show
that the density of states and the entropy in such a framework are
less than the corresponding quantities in Einstein's Special
Relativity (SR) theory. This result can be interpreted as a
consequence of the deformed Lorentz symmetry present in the
particular model we have considered.

\new \n {\section{\textbf{Introduction}}}

\vspace{.5cm} Any description of Quantum Gravity suggests a
smallest (but finite) length scale $l$ (or a finite upper bound of
energy $\kappa$), which of course should be observer independent.
The natural candidate for this is the Planck Length (or the Planck
Energy). But this proposition obviously contradicts the principles
of Einstein's Special Relativity (SR) Theory, as in SR, the length
or the mass (or energy) of an object varies for different
observers. Thus we need an extension of SR theory where along with
the velocity of light, another observer-independent quantity, a
fundamental length-scale exists. As a consequence, there must be
some modifications of SR theory in the high energy (Planck energy) regime.\\
As a possible solution, a new theory (DSR Theory) was first
proposed by Amelino-Camelia \cite{ac1}. Another model, perhaps
simpler, was given by Amelino-Camelia \cite{ac2} and by Magueijo
and Smolin \cite{mag1} (for discussion and review, see
\cite{acr,kowglik1} and references therein). As said earlier, in
these theories, there are two invariant quantities, $c$, the
velocity of light and $\ka$, an upper limit of energy. But for
consistent inclusion of this second invariant quantity along with
the other principles of SR theory, the well known dispersion
relation (or mass-shell condition) for a particle \be E^2 - p^2 =
m^2 \ee has to be modified as: \be E^2 - p^2 = m^2 \left(1 -
\frac{E}{\ka}\right)^2. \label{mdr} \ee Here $E$ and $p$ are
respectively the energy and the magnitude of the three-momentum of
the particle, $m$ is the mass of the particle and we have taken
$c=1$. We refer this model as the Magueijo-Smolin (MS) model.\\
In earlier work \cite{dasghosh}, we considered a particular
dispersion relation as in \cite{mag1}. Then we derived an
expression for the energy-momentum tensor for a perfect fluid and
studied dynamics of the perfect fluid with this modified
expression. Due to the presence of the invariant energy scale, our
derivation of the energy-momentum tensor was subtle where
nonlinear representation of Lorentz transformations played an
essential role. In this work, we adopt the same scheme and
consider (\ref{mdr}) as our fundamental equation. Then we go on to
study the thermodynamic properties of an ideal photon gas using
the methods of conventional statistical mechanics, but generalized
to be applicable in a theory where
an invariant energy scale is present.\\
We have arranged this paper as follows:~~In section 2, we discuss
about the modified dispersion relation. In the next section we
derive the expression for the density of states and the important
expression of the partition function. The derivation of the
expression for partition function is the most crucial result of
our work. In section 4, we go on to study the thermodynamic
properties of photon gas using this partition function. In
particular, we evaluated analytic expressions for the pressure,
equation of state, internal energy, entropy and specific heat of
the photon gas. We also show the comparisons between the
thermodynamic variables in the MS model and in the usual SR
scenario. Further, we see that the density of states as well as
the entropy decreases in the MS model as compared to that in the
SR framework. This happens due to the deformation in Lorentz
symmetry in the theory where an invariant energy scale is present.
It is another major result of our work. Finally, we conclude
summarizing our results and discuss some of the future prospects
in this regard.

\vspace{1cm} \n {\section{\textbf{Modified Dispersion Relation}}

\vspace{.5cm} We choose a particular modified dispersion relation
as given in \cite{mag1,dasghosh} $$ E^2 - p^2 = m^2 \left(1 -
\frac{E}{\ka}\right)^2. $$ Thermodynamic properties for photon gas
with a different dispersion relation has been studied in
\cite{camacho1}. Also, thermodynamics of bosons and fermions with
another modified dispersion relation and its cosmological and
astrophysical implications has been observed in \cite{magcos,
bertolami}. But these two modified dispersion relations appear
from a phenomenological point of view whereas the dispersion
relation (\ref{mdr}) has a more theoretical motivation which we
discuss below in some details. \\
It was shown in \cite{sghosh} that existence of an invariant
length scale in the theory is consistent with a non-commutative
(NC) phase space ($\ka$-Minkowski spacetime) such that the usual
canonical Poisson brackets between the phase space variables are
modified. Also, the linear Lorentz transformations (LT) are
replaced by non-linear $\ka$-Lorentz transformations ($\ka$-LT)
\cite{bruno, sghosh}. But still Lorentz algebra is intact in the
theory. As a result, we have the $\ka$-LT invariant modified
dispersion relation (\ref{mdr}) as: \be \{J_{\mu \nu} ,
\frac{p^2}{\left(1 - \frac{E}{\ka} \right)^2}\} = 0.
\label{invariant} \ee The angular momentum $J_{\mu \nu}$ is
defined as in \cite{sghosh} $$ J_{\mu \nu} = x_\mu p_\nu - x_\nu
p_\mu$$ where $x$ and $p$ are the phase space variables. Due to
the nontrivial expression for the dispersion relation (\ref{mdr}),
firstly it was supposed that the velocity of photon $ c = \frac{d
E}{d p} $ have to be energy dependent. But it was shown in
\cite{hoss} that a modified dispersion relation does not
necessarily imply a varying (energy dependent) velocity of light.
Thus, though the above two models (\cite{camacho1} and
\cite{magcos, bertolami}) admit a varying speed of light, in case
of MS model, for photons ($ m = 0$) the dispersion relation
(\ref{mdr}) is the same as in SR theory. Also the speed of light
$c$ is an invariant quantity in the MS model \cite{mag1, sghosh,
dasghosh}. Thus the MS model considered in \cite{mag1, sghosh} has
a more theoretical motivation and it can be developed starting
from the NC phase space variables \cite{sghosh} whereas the models
considered in \cite{camacho1, magcos, bertolami} are
phenomenological in nature and as far as we know, there is no
fundamental phase space structures to describe these models.\\
Another interesting fact is that both the models described in
\cite{camacho1} and in \cite{magcos, bertolami} have no finite
upper bound of energy of the photons though they have a momentum
upper bound. But, as stated earlier, in the MS case, though the
dispersion relation for the photons is unchanged, there is a
finite upper bound of photon energy which is the Planck energy
$\ka$. One can readily check that this is an invariant quantity by
using the $\ka$-Lorentz transformation law
for the energy \cite{bruno, sghosh}. \\
One more thing must be clarified here. In case of the models
(\cite{camacho1} and \cite{magcos, bertolami}), clearly the
Lorentz symmetry was broken and as a result, the number of
microstates and hence the entropy increases as compared to the
Lorentz symmetric SR theory. On the other hand, we are dealing
with a different scenario where the Lorentz symmetry is not broken
as Lorentz algebra between the phase space variables is intact. In
fact, the framework we describe here still satisfies the basic
postulates of Einstein's SR theory; moreover it possesses another
observer-independent quantity \cite{acr}. Thus it seems that
Lorentz symmetry is further restricted in this MS model. As a
result of this, we expect to have a less number of microstates and
less entropy in the Ms model. As we will show later in our
explicit calculations, this expected result is correct.\\
As we have said earlier, the modified dispersion relation
(\ref{mdr}) in case of the photons (massless particles) does not
change from the usual SR scenario. Thus, for the photons, the
dispersion relation now becomes \be p = E.
\label{photondispersion} \ee

\vspace{1cm} \n {\section{\textbf{Partition Function for Photon
Gas}}

\vspace{.5cm} To study the thermodynamic behavior of photon gas,
we have to find out an expression for the partition function
first, as it relates the microscopic properties with the
thermodynamic (macroscopic) behavior of a physical system
\cite{pathria, greiner}, which we do in this section.\\

\n {\subsection{Number of states:}} \vspace{.2cm}

We consider a box containing photon gas. Following the standard
procedure as given in \cite{pathria, greiner}, we consider a
continuous spectrum of momentum instead of quantizing it. The
number of microstates available to the system ($\sum$) in the
position range from $r$ to $r+dr$ and in the momentum range from
$p$ to $p+dp$ is given by \cite{pathria, greiner} \be \sum =
\frac{1}{h^3} \int \int d^3 \vec{r} d^3 \vec{p} \label{state1} \ee
where $h$ is the phase space volume of a single lattice and $$\int
\int d^3 \vec{r} d^3 \vec{p}$$ is the total phase space volume
available to the system. \\
It should be mentioned here that in case of SR theory, the
quantities $E d^3 x$ and $\frac{d^3 p}{E}$ are invariant under the
Lorentz transformations and hence the phase space volume element
$d^3 x d^3 p$ is a Lorentz invariant quantity \cite{misner}. The
nonlinear $\ka$-Lorentz transformations \cite{bruno, sghosh} are
explicitly given by: \be t' = \alpha \gamma (t - v x)~~,~~x' =
\alpha \gamma (x - v t)~~,~~y' = \alpha y~~,~~z' = \alpha z $$$$
E' = \frac{\gamma (E - v p_x)}{\alpha}~~,~~p'_x = \frac{\gamma
(p_x - v E)}{\alpha}~~,~~p'_y = \frac{p_y}{\alpha}~~,~~p'_z =
\frac{p_z}{\alpha}. \label{kalt} \ee The prime over a quantity
denotes the corresponding quantity in the boosted frame and
$\alpha = 1 + \frac{1}{\kappa}((\gamma - 1) E - v \gamma p_x)$. We
have considered the three-momentum to be of the usual form: $
\vec{p} = (v E, 0, 0)$ and $\gamma = \frac{1}{\sqrt{1-v^2}}$ where
$v$ is the velocity of the boosted frame. In case of our model,
the phase space volume element $d^3 x d^3 p$ is invariant under
the $\kappa$-Lorentz transformations (\ref{kalt}) as: \be d^3 x'
d^3 p' = \alpha^3 \gamma~d^3 x~ \frac{\gamma}{\alpha^3} \left(1 -
\frac{v p_x}{E}\right)d^3 p = d^3 x d^3 p. \label{invphsp} \ee It
is interesting to note that the factor $\alpha$ arising from the
nonlinear $\kappa$-Lorentz transformation finally cancels out in
(\ref{invphsp}). Strictly speaking, to derive (\ref{invphsp}) we
should consider the effect coming from variation of $\alpha$. But
we have omitted the term $d \alpha$ in (\ref{invphsp}) as it is a
dynamical effect and may not be relevant for the free particle
case as considered here.

If the volume of the box is considered to be $V$, the number of
microstates can be written in the following form using the
spherical polar coordinates \cite{pathria, greiner} \be \sum =
\frac{4 \pi V}{h^3} \int_0^{\infty} E^2 d E \label{state2}. \ee We
used the dispersion relation $ p=E$ to change the integration
variable to $E$. Then considering the fact that we have an finite
upper limit of energy ($\ka$), we obtain the number of
microstates: \be {\tilde{\sum}} = \frac{4 \pi V}{h^3} \int_0^{\ka}
E^2 d E \label{state}, \ee where $ \sim $ on a quantity represents
the corresponding quantity in the model we have considered. It is
obvious from the expressions (\ref{state2}) and (\ref{state}) that
the available number of microstates to the system is less than
that in the SR theory, as the energy spectrum of a particle in SR
theory can go all the way up till infinity.
This result agrees with our expectation stated earlier. \\

\n {\subsection{Partition function:}} \vspace{.2cm}

It is very crucial to get an expression for the partition function
as all the thermodynamic properties can be thoroughly studied
using the knowledge about the partition function. The single
particle partition function $Z_1(T, V)$ is defined as
\cite{greiner} \be Z_1(T, V) = \frac{4 \pi V}{h^3} \int_0^{\infty}
p^2 e^{-\beta E} d p, \label{partition1} \ee where
$\beta=\frac{1}{k_B T}$, $k_B$ is the
Boltzman constant and $T$ is the temperature of the particle.\\
For the MS model, the single particle partition function
$\tilde{Z}_1(T, V)$ is defined as \be \tilde{Z}_1(T, V) = \frac{4
\pi V}{h^3} \int_0^{\ka} p^2 e^{-\beta E} d p.
\label{partition1dsr} \ee In the limit $\ka \rightarrow \infty$,
we should get back normal SR theory results. \\
It should be noted that in the MS model which we have considered,
the photon dispersion relation is not modified at all. But still
there is modification in the partition function
(\ref{partition1dsr}) due to the presence of an energy upper bound
of particles ($\ka$) in the theory. So the upper limit of
integration is $\ka$ in (\ref{partition1dsr}) whereas in the
normal SR theory expression (\ref{partition1}), the upper limit of
integration is $\infty$ since there is no upper bound of energy in
the SR theory. In all the models \cite{camacho1, magcos,
bertolami}, though the upper limit of energy is infinity as in SR
theory, these models differ
due to the different dispersion relations. \\
Using the dispersion relation for photons ($E = p$) and using the
standard table and formulae for integrals \cite{gradstein}, we
finally have an analytical expression of the single particle
partition function \be \tilde{Z}_1(T, V) = \frac{4 \pi V}{h^3}
\int_0^{\ka} E^2 e^{-\beta E} d E = \frac{4 \pi V}{h^3}
\left[\frac{2}{\beta^3} - \frac{e^{-\beta \ka}}{\beta^3}(2+\beta
\ka (2+\beta \ka))\right]. \label{part1} \ee Thus the partition
function for a $N$-particle system $\tilde{Z}_N(T, V)$ is given by
\be \tilde{Z}_N(T, V) = \frac{1}{N!} [\tilde{Z}_1(T, V)]^N
\label{partn} \ee where we have considered classical
(Maxwell-Boltzman) statistics along with the Gibb's factor. As we
get the expression for the partition function, now we go on to
study various thermodynamic properties of the photon gas in our
model. It should be noted that as $\ka \rightarrow \infty$, this
partition function coincides with the partition function in SR
theory and thus all of our results coincides with the usual SR
case in this limit. \vspace{1cm}

\n {\section{\textbf{Thermodynamic Properties of Photon Gas}}}

\vspace{.5cm}

With the expression for the partition function in our hand, now we
go on to study various thermodynamic properties of photon gas in a
theory where an observer-independent fundamental energy scale is present.\\

\n {\subsection{Free energy:}} \vspace{.2cm}

We use Stirling's formula for $ ln [N!]$ \cite{greiner} $$ ln [N!]
\approx N~ln[N] - N$$ in the expression for partition function
(\ref{partn}) to obtain the free energy $\tilde{F}$ of the system
\be \tilde{F} = -k_B T~ln [\tilde{Z}_N(T, V)] $$$$ = - N k_B T
\left[ 1 + ln \left[\frac{4 \pi V}{N}\left(\frac{k_B
T}{h}\right)^3 \left\{2 - e^{- \frac{\ka}{k_B T}}\left(2 +
\frac{\ka}{k_B T}\left(2 + \frac{\ka}{k_B
T}\right)\right)\right\}\right] \right]. \label{free} \ee In the
limit $\ka \rightarrow \infty$, the terms containing $\ka$
vanishes and we get back normal SR theory result:
$$ F = N k_B T. $$ \vspace{.2cm}

\n {\subsection{Pressure:}} \vspace{.2cm}

From the expression for free energy (\ref{free}) we can readily
obtain the pressure $\tilde{P}$ of photon gas in our considered
model as \cite{greiner} \be \tilde{P} = - \left(\frac{\partial
F}{\partial V}\right)_{T, N} = \frac{N k_B T}{V}. \label{pressure}
\ee Thus, we have the same equation of state
$$ P V = N k_B T$$ as in SR theory. \vspace{.2cm}

\n {\subsection{Entropy:}} \vspace{.2cm}

As we have the expression for free energy (\ref{free}), also we
can evaluate the entropy $\tilde{S}$ of the system from the
following relation \cite{greiner} $$ \tilde{S} =
-\left(\frac{\partial F}{\partial T}\right)_{V, N}. $$ The
expression for entropy takes the following form \be \tilde{S} = N
k \left[4 + ln \left[\frac{4 \pi V}{N}\left(\frac{k_B
T}{h}\right)^3 \left\{2 - e^{- \frac{\ka}{k_B T}}\left(2 +
\frac{\ka}{k_B T}\left(2 + \frac{\ka}{k_B T}\right) \right)
\right\}\right] \right. $$$$ \left. - \frac{\ka^3}{2 k_B^3 T^3
e^{\frac{\ka}{k_B T}} - \left(2 k_B^3 T^3 + 2 k_B^2 T^2 \ka + k_B
T \ka^2\right)}\right]. \label{entropy} \ee The terms containing
$\ka$ are the modifications from the SR theory expression of
entropy \cite{greiner}. As in the earlier expressions, in the
limit $\ka \rightarrow \infty$ the terms containing $\ka$ vanish
and we get back the SR theory result: \be S = N k_B \left[4 + ln
\left[\frac{8 \pi V}{N}\left(\frac{k_B T}{h}\right)^3 \right]
\right]. \ee We plot the entropy $S$ against $T$ both for the
model we considered and for SR theory to study the deviation of
entropy in the two models.

\begin{figure}[htb]
{\centerline{\includegraphics[width=9cm, height=5cm] {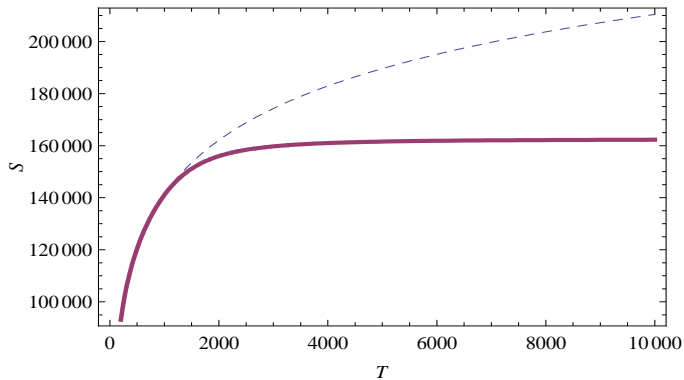}}}
\caption{{\it{Plot of entropy of photon $S$ against temperature
$T$ for both in the SR theory and in our case; the dashed line
corresponds to the SR theory result and the thick line represents
the corresponding quantity in our result. We have used the Planck
units and the corresponding parameters take the following values $
\ka = 10000, k_B = 1, N = 10000, V = .01, h = 1 $ in this plot as
well as in all other plots in the paper. In this scale, $T=10000$
is the Planck temperature.}}} \label{fig1}
\end{figure}

In Figure 1, we have plotted entropy against temperature for both
the case of our invariant energy scale scenario and normal SR
theory.  It is clearly observable from the plot that the entropy
grows at a much slower rate in case of our result than in the SR
theory and as temperature increases, the entropy in our considered
model deviates more from the entropy in the SR theory. This result
matches with our earlier expectation considering the underlying
symmetry of the theory that the entropy in the MS model should be
less than the entropy in SR theory.\\
It is well known that the total number of microstates available to
a system is a direct measure of the entropy for that system.
Therefore our result merely reflects the fact that due to the
existence of an energy upper bound $\ka$, the number of
microstates gradually saturates to some finite value near Planck
scale. \vspace{.2cm}

\n {\subsection{Internal energy:}} \vspace{.2cm}

We expect modification in the expression of the internal energy
$U$ for photon gas in the MS model as the expression of entropy is
modified and internal energy is related to the entropy as follows:
$$ U = F + T S. $$ In the usual SR scenario, the explicit
expression for internal energy is given by \be U = 3 N k_B T.
\label{internalsr} \ee But in the MS scenario we considered, the
expression for internal energy ($\tilde{U}$) of photon gas takes
the following form \be \tilde{U} = N k_B T \left[3 - \frac{ \ka^3
e^{- \frac{\ka}{k_B T}}}{2 k_B^3 T^3 - e^{- \frac{\ka}{k_B T}}(2
k_B^3 T^3 + 2 \ka k_B^2 T^2 + \ka^2 k_B T)}\right].
\label{internaldsr} \ee It is easy to see from the expression of
internal energy (\ref{internaldsr}) that we get back the usual SR
theory expression in the limit $\ka \rightarrow \infty$. As in the
case of entropy, here we also plot internal energy against
temperature for both the SR and MS case.

\begin{figure}[htb]
{\centerline{\includegraphics[width=9cm, height=5cm] {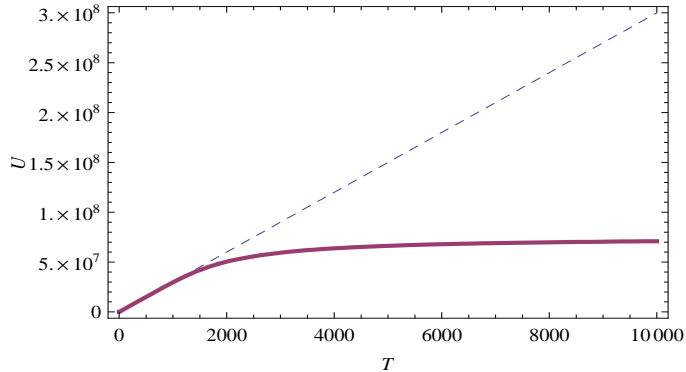}}}
\caption{{\it{Plot of internal energy of photon $U$ against temperature
$T$ for both in the SR theory and MS scenario; the dashed line
corresponds to the SR theory result and the thick line represents
the quantity in the MS model we considered here.}}} \label{fig2}
\end{figure}

In Figure 2, we plotted internal energy of photon gas against its
temperature for both the case of MS model and SR theory. The
expression in SR theory (\ref{internalsr}) tells us that internal
energy depends linearly on the temperature and this is supported
from the plot. But from the expression of internal energy in the
MS model (\ref{internaldsr}) it is clearly observed that the
relation of internal energy with temperature is not linear at all.
Also one can easily check that the value of internal energy (for a
particular temperature) in the MS model (\ref{internaldsr}) is
always less than its value (for the same temperature) in the SR
theory (\ref{internalsr}). This is very clear from the plot as the
curve for the MS model always lies below the straight line which
corresponds the SR theory result. \\
Since the internal energy $U$ of photon gas becomes saturated
after a certain temperature in case of the MS model, it is
tempting to point out that probably our results are moving towards
the right direction related to the ``Soccer Ball Problem" that
plagues multi-particle description in the framework of DSR. The
problem lies in the fact that if we apply linear addition rule for
momenta/energies of many sub-Planck energy particles we may end up
with a multi-particle state, such as a soccer ball, whose total
energy is greater than the Planck energy which is forbidden in the
DSR theory. For further discussion about the ``Soccer Ball
Problem" see \cite{maghoss}. \vspace{.2cm}

\n {\subsection{Pressure-energy density relation:}} \vspace{.2cm}

Though internal energy $U$ of a physical system is not directly
measurable, still we can detect the effect of it through other
thermodynamic quantities, such as the relation between pressure
$P$ and energy density $\rho$ of that system. Energy density of a
system $\rho$ is defined as $$ \rho = U/V $$ where $U$ is the
internal energy of the system and $V$ is the volume occupied by
the system. As the expression for internal energy $U$ is modified
in the MS model, we also expect modifications in the expression
for the energy density $\rho$. The modified relation between
pressure $\tilde{P}$ and energy density $\tilde{\rho}$ is given
by: \be \tilde{P} = \frac{1}{3} \tilde{\rho} + \frac{1}{3} \frac{
N k_B T \ka^3 e^{-\frac{\ka}{k_B T}}}{2 V k_B^3 T^3 - V
e^{-\frac{\ka}{k_B T}}(2 k_B^3 T^3 + 2 \ka k_B^2 T^2 + \ka^2 k_B
T)}. \label{prenergy} \ee For $\ka \rightarrow \infty$, we get
back the usual pressure-energy density relation in SR theory: $$ P
= \frac{1}{3} \rho. $$ It should be pointed out that in our
earlier work \cite{dasghosh}, it was shown that in the
ultra-relativistic regime (for photons), the relation $P =
\frac{1}{3} \rho$ remains unaffected. But here we have a
modification in this pressure-energy density relation
(\ref{prenergy}). In \cite{dasghosh}, we obtained the result
considering some simplified assumptions. But in this work, we
start with the partition function and apply the methods of
Statistical Mechanics (which naturally deals with multi-particle
systems). So we do not really have to consider any strong
assumptions here. \vspace{.2cm}

\n {\subsection{Specific heat:}} \vspace{.2cm}

There is another thermodynamic parameter, specific heat ($C_V$),
through which we can observe the modifications in the expression
for internal energy. Specific heat $C_V$ is defined as $$ C_V =
\left(\frac{\partial U}{\partial T}\right)_V. $$ For the MS model
we considered here, explicit calculation yields the following
result \be \tilde{C}_V = 3 N k_B - \frac{2 N k_B \ka^3 \left(k_B^2
T^2 \ka(1 + e^{\frac{\ka}{k_B T}}) + 2 k_B^3 T^3 (1 -
e^{\frac{\ka}{k_B T}})\right)}{\left(2 k_B^3 T^3
(e^{\frac{\ka}{k_B T}} - 1)-2 k_B^2 T^2 \ka - k_B T \ka^2
\right)^2}. \label{spheat} \ee After doing a bit of algebra, one
can check from the above expression (\ref{spheat}) that specific
heat calculated from the MS model is always less than the value
calculated from usual SR theory. Also, when $ \ka \rightarrow
\infty$, we obtain $$ C_V = 3 N k_B $$ which is the usual specific
heat for photon as calculated in SR theory.

\begin{figure}[htb]
{\centerline{\includegraphics[width=9cm, height=5cm] {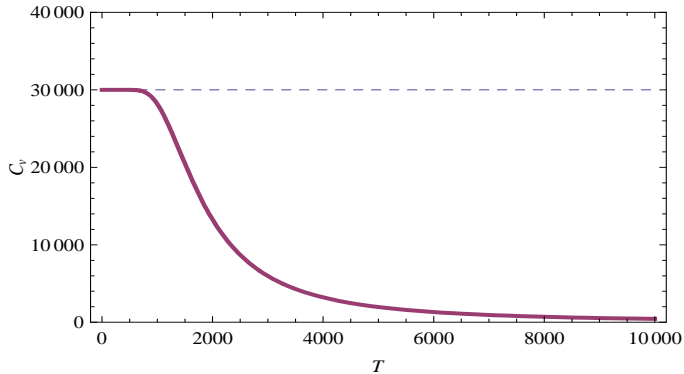}}}
\caption{{\it{Plot of specific heat of photon $C_V$ against temperature
$T$ for both in the SR theory and MS scenario; the dashed line
corresponds to the SR theory result and the thick line represents
the quantity in the MS model we considered here.}}} \label{fig3}
\end{figure}

In Figure 3, we have plotted the specific heat $C_V$ against
temperature $T$. It is clear from the plot that in the case of the
MS model, the specific heat $C_V$ asymptotically decreases to zero
suggesting that photon gas has reached its temperature ceiling
which is the Planck temperature.

\n {\section{\textbf{Conclusion and Future Prospects}}}
\vspace{.2cm}

We consider the modified dispersion relation as given in the MS
model \cite{mag1}. We explicitly show that for photons, the number
of microstates available to a macrostate is less in the MS model
than in the usual SR scenario. We stress that it happens since
Lorentz symmetry is not broken in this model. But due do the
presence of an invariant energy upper bound in this theory,
microstates can avail energies only up to a finite cut-off whereas
in SR theory, microstates can attain energies up to infinity.
Thus, quite naturally, the number of microstates in this MS model
is less than that in SR theory. \\
The most significant result of our work is the derivation of
$N$-particle partition function in the MS model. Due to the
presence of the deformed dispersion relation, this task becomes
highly non-trivial. However, for photons, we find out an analytic
expression for the partition function. Once we have the partition
function in our hand, we evaluate other various thermodynamic
parameters of photon gas such as the free energy, pressure,
entropy, internal energy, specific heat for the MS model and
compare them with the known results of SR theory. \\
As a consequence of deformed Lorentz symmetry, the entropy in the
MS model is also less than that in the SR scenario. We show this
behavior analytically and graphically. Also the internal energy is
modified in case of the MS model and as a consequence the
expression for the specific heat is also modified.\\
Though highly non-trivial, one can similarly study the behavior of
an ideal gas using this modified dispersion relation. Also one can
study behavior of fermion gas in this MS model. There might be
some modifications in the Fermi energy level which can modify the
Chandrasekhar mass limit for the white dwarf stars
\cite{bertolami}. Thus astrophysical phenomena in an MS
framework is another issue remains to be addressed.\\
Further, as we have the expression for energy-momentum tensor, one
can study the cosmological aspects of MS model using the Friedmann
equations. But this requires idea about the geometry sector
(precisely the metric $g_{\mu \nu}$ and hence Einstein tensor
$G_{\mu \nu}$) which is till unknown in the context of MS model.
This still remains another open issue to be further studied.

It is noteworthy to mention here that ``bouncing" loop quantum
cosmology theories (for example see \cite{bojowald} and references
therein) entails some modifications to the geometry of spacetime
which in turn effectively puts a bound on the curvature avoiding
the big bang singularity. However, for these ``bouncing" models,
the perturbation technique cannot be done as at the point of
curvature saturation, the energy density of the cosmic fluid
diverges. So it is unclear how to construct the matter part of
Einstein equation. One alternative to avoid the big bang
singularity is the inflation theory where the perturbation method
can also be applied. On the other hand, in our model, the energy
density of the cosmic fluid saturates to the Planck energy which
is a finite real quantity. Possibly a combination of the model
considered in this paper along with the ``bouncing" loop quantum
cosmology can successfully describe a situation where big bang
singularity can be avoided. \vspace{.5cm}

\n {\textbf{Acknowledgements}} \vspace{.2cm}

We would like to thank Prof. Subir Ghosh for helpful discussions.

\ed